# BIOMERO: BioImage analysis in OMERO

Torec T. Luik[1*], Rodrigo Rosas-Bertolini[1], Eric A.J. Reits[1], Ron A. Hoebe[1¶], Przemek M. Krawczyk[1,2¶*]

[1] Amsterdam UMC, Department of Medical Biology, Amsterdam, The Netherlands

[2] lead contact

* Corresponding authors

E-mail: p.krawczyk@amsterdamumc.nl (PK); t.t.luik@amsterdamumc.nl (TL)

¶ These authors contributed equally to this work.

## Summary

In the rapidly evolving field of bioimaging, the integration and orchestration of Findable, Accessible, Interoperable, and Reusable (FAIR) image analysis workflows remains a challenge. We introduce BIOMERO, a bridge connecting OMERO, a renowned bioimaging data management platform, FAIR workflows and high-performance computing (HPC) environments. BIOMERO, featuring our open-source Python library "OMERO Slurm Client", facilitates seamless execution of FAIR workflows, particularly for large datasets from High Content or High Throughput Screening. BIOMERO empowers researchers by eliminating the need for specialized knowledge, enabling scalable image processing directly from OMERO. BIOMERO notably supports the sharing and utilization of FAIR workflows between OMERO, Cytomine/BIAFLOWS, and other bioimaging communities. BIOMERO will promote the widespread adoption of FAIR workflows, emphasizing reusability, across the realm of bioimaging research. Its user-friendly interface will empower users, including those without technical expertise, to seamlessly apply these workflows to their datasets, democratizing the utilization of AI by the broader research community.

## Introduction

In the realm of modern bioimaging, efficient management and analysis of large image datasets stand as pivotal challenges. Open-source bioimaging platform OMERO[1] excels in image data management, supporting all major proprietary microscopy data and metadata formats, and allowing users to view and annotate their images in a web browser. OMERO is frequently used by Core Facilities in the life sciences to store and manage microscopy data for their users. While OMERO supports Python scripts, it currently lacks native capabilities for conducting (remote) data analysis. The prevailing approach involves exporting data from OMERO to user-installed applications, for example FIJI[2] or QuPath[3], resulting in the analysis computation occurring on the user's end rather than within the integrated data management suite. However, utilizing the simple Python script capabilities on the OMERO server also falls short of addressing crucial aspects: the lack of resource management, such as CPU, GPU and memory utilization, due to the absence of a robust queue mechanism within OMERO, and the inherent clash of hardware requirements between data management, image viewing, and image analysis. This not only poses a question about scalability of compute power for larger images or datasets, but also about reproducibility of the image analysis conducted using such applications, especially when considering high-throughput applications like High Content or High Throughput Screening (HCS/HTS). Hence, we identify two major challenges: first the need for cohesive integration and orchestration of FAIR workflows from OMERO. Second, the ability to harness High-Performance Computing (HPC) resources effectively for analyzing big image datasets, especially in HCS/HTS contexts.

The landscape of existing solutions consists of several tools addressing specific aspects of bioimaging challenges:

1. First and foremost, Slurm Workload Manager, often simply referred to as Slurm, is a job scheduler available on many of the world's high-performance clusters[4], which enables efficient distribution of workflow workloads over multiple computers. To use it, however, one needs to login to a specific server using a command-line interface, download all the required data and tools and then run jobs scripts with a specific hardware configuration, which requires specialist DevOps knowledge not prevalent among researchers.
2. Cytomine[5] specializes in managing whole-slide imaging (WSI) data and digital pathology but lacks support for HCS/HTS and is less used in general microscopy Core Facilities. Cytomine does, however, excel in the scalable execution of modular Cytomine Apps (workflows), integrated into Cytomine UI, by executing them on an HPC cluster with Slurm while abstracting these operations away from the user.

3. BIAFLOWS[6] is an extension of Cytomine, funded and developed by NEUBIAS to support FAIR workflows across bioimaging. It particularly enables comparison of workflow results and increases interoperability of such workflows, by also enabling execution outside of the Cytomine environment. Neither Cytomine nor BIAFLOWS, however, offer integration with OMERO, making it very difficult to execute their workflows on OMERO-stored data.

4. "ObiWan-Microbi"[7] is a recent tool focusing on Microbe colony and segmentation with the ability to execute workflows from its own model zoo and offload computations to another server. While it integrates segmentation features with OMERO, it falls short in accommodating a diverse range of FAIR containers, such as BIAFLOWS apps. Notably, it lacks support for container technology and efficient offloading to HPC clusters, particularly lacking SLURM/HPC queueing capabilities.

5. OMERO Plus[8], a proprietary version by Glencoe Software, provides extra computational capabilities integrated with OMERO, including the OMERO-CellProfiler Connector[9] for remote execution of CellProfiler pipelines via OMERO clients on different HPC systems and cloud platforms. However, it also lacks support for more generic FAIR workflows and, more importantly, is not free and not publicly available to the bioimaging community.

6. The Fractal Analytics Platform[10], a tool still in its early stages and active development, focuses on workflow computations on ZARR files only. However, it lacks direct integration with source data and UI, and requires managing a separate server/client architecture next to OMERO. Moreover, it also does not support generic FAIR containerized workflows, like BIAFLOWS apps.

To enable both execution of FAIR workflows and HPC support in an OMERO environment, we present BIOMERO, BioImage analysis in OMERO. BIOMERO is an open-source framework aimed at integrating bioimaging data storage and management with high-throughput, FAIR image analysis pipelines. BIOMERO consists of the OMERO Slurm Client[11] (OSC) Python library and tools for its integration with the OMERO user interface. BIOMERO extends the capabilities of OMERO, providing efficient access to High-Performance Computing (HPC) resources and enabling scalable, FAIR workflows in the bioimaging field. By supporting BIAFLOWS workflows, BIOMERO fills a critical gap in existing infrastructure, facilitating the execution of intricate bioimaging analyses and unifying OMERO and the Cytomine/BIAFLOWS platforms under a common umbrella of FAIR workflows.

# Results

## Architecture and design

BIOMERO is a framework that connects OMERO and HPC, redefining the architecture of bioimaging workflows for enhanced scalability and promoting adherence of such workflows to FAIR principles. BIOMERO is a combination of the OMERO Slurm Client (OSC) python library, as an API to Slurm, and a set of OMERO HPC core scripts that allow user interaction from within the OMERO web interface. To make use of BIOMERO, an existing OMERO server and an existing Slurm HPC cluster need to be available.

Figure 1 shows an overview of the BIOMERO architecture. At the core there is the OMERO Slurm Client (OSC), a Python package installed by the administrator on the OMERO server, providing an Application Programming Interface (API) to Slurm on the HPC cluster. This interface can be accessed from OMERO.scripts, which regular users can run from OMERO web interface (OMERO.web). Through this Python API, the scripts can then command BIOMERO (A) to send OMERO data to HPC and start analysis jobs on the HPC cluster (B), poll their progress (C), and retrieve results to store back in OMERO (D). With BIOMERO, users don't need to be versed in Slurm commands, data transfer procedures, or the specifics of Docker container and application codes. BIOMERO abstracts most of the data and workflow management, ensuring a wider range of users can leverage the distributed compute power directly from the OMERO web interface to run predefined workflows (e.g. segmentation) on their data.

## OMERO Slurm Client

The OMERO Slurm Client (OSC) Python library, is released on GitHub with the permissive Apache 2.0 license and published on PyPI for easy installation on OMERO servers. The package contains a single class, the *SlurmClient*, which serves as the API to a connected HPC cluster running Slurm. It builds upon the Python library Fabric[12], which enables running commands on remote machines through a secure SSH connection; this allows OSC to connect OMERO server to any remote Slurm cluster that can be accessed with SSH, whether in the cloud, locally or at a centralized data center. OSC also adds an abstraction layer via dedicated Python functions, covering all major parts of the workflow lifecycle.

The library is configured by '*slurm-config.ini*', where the OMERO administrator can describe what workflows should be available to end-users and where to find them online (GitHub repository and version), as well as the SSH settings for the target Slurm cluster. Once configured, the *SlurmClient* class can connect to the existing Slurm cluster and provides a function to automatically configure a managed folder structure (see Figure S1), including downloading the container images for the workflows.

Once the Slurm environment is ready, it can execute the preconfigured workflows as jobs. To run a workflow as a job, Slurm requires a job script specifying the required hardware, container image and job-related parameters such as input folder and workflow settings. The Slurm job script will by default be generated by OSC from the workflow's metadata (the descriptor file) stored on GitHub, but can either be adjusted partially or entirely pulled from a GitHub repository (both configurable in the *slurm-config.ini* file). A full list of commands and functions available from the OSC can be found in the latest documentation[13].

## OMERO HPC core scripts and web interface

The second part of BIOMERO entails a set of OMERO scripts. We enhance OMERO's scripting capabilities, providing an intuitive way for users to interact with and execute configured workflows directly from the OMERO web interface. These OMERO scripts use the OSC library, described previously, to run workflows as Slurm jobs, poll the jobs and, most importantly, get data back and forth from OMERO to the Slurm cluster. We have provided example implementations for running workflows in OMERO web interface (OMERO.web) through several example scripts on GitHub[14] This system is highly extensible and modular system, and we encourage the community and OMERO administrators to create their own scripts for their specific workflow or User Interface needs.

OMERO HPC core scripts that we have provided are shown below.

- `SLURM Init Environment`, which will take care of creating the setup in Figure S1 based on the configuration file;
- `SLURM Run Workflow Batched`, which allows users to run any of the configured workflows on select input images, datasets or plates. It creates a job for every batch of input images to run in parallel on Slurm, it will poll execution of these jobs and only return when all jobs are finished. It uses the parameters read from the workflow descriptor to create UI elements for OMERO web
    - `SLURM Run Workflow`, which is called by *Batched* per batch, to run any of the configured workflows as one Slurm job on all provided input images, datasets or plates. It calls the scripts `SLURM Image Transfer` to transfer data from OMERO and, when the Slurm job has finished, `SLURM Get Results` to get the results back into OMERO
        - `SLURM Image Transfer`, an adaptation of OME's `Batch Image Export` script[15], to export images from OMERO as ZARR and transfer them to the HPC using secure copy protocol (SCP)[16] . See Note S2 for a discussion on using ZARR here.
        - `SLURM Get Results` can upload a folder from HPC back into OMERO, either as images in a dataset, as attachments to selected input files, or as a generic zip attachment to a parent project.

# Executing FAIR workflows from OMERO

BIOMERO advocates for FAIR principles in bioimaging research through the execution of BIAFLOWS workflows. Leveraging publication on GitHub and DockerHub, metadata descriptors, and Docker or Singularity containerization, BIAFLOWS workflows facilitate the adoption of Findable, Accessible, Interoperable, and Reusable practices. BIAFLOWS workflows are essentially derived from Cytomine Apps, but offer enhanced FAIR attributes, especially in their capacity to be executed independently of Cytomine, emphasizing increased interoperability. In conclusion, BIOMERO seamlessly supports BIAFLOWS workflows, and with minor adjustments it can also accommodate Cytomine Apps or any FAIR workflow with other metadata descriptors.

In Figure 2, we show the files required to make any workflow compatible with the BIOMERO framework. Any executable code that can run headless (i.e. without a graphical user interface) from the command-line (f.e. a Fiji Macro, a compiled Matlab script, a Python script, an R script, a .exe) should be packaged up by adding 3 files: a Dockerfile to define the environment in which the executable code can run (f.e. OS, Python version, Python libraries, Fiji plugins), a descriptor JSON file that defines the metadata of the code (e.g. command-line parameters, what values they take and their descriptions), and a short Python script that wraps the executable code (running it as a Python subprocess) and provides it with the correctly preprocessed input and output data. These 4 files should be published online on a public code versioning repository (e.g. on GitHub). With these files in place a Docker container image can be built, which will allow the executable code to be executed from any platform with container software like Docker, Podman, or Singularity installed. However, to ensure reproducibility, the built container image should also be published and versioned in a public container registry like DockerHub. This process can be automated, via GitHub Actions for instance, to keep versions of GitHub and DockerHub the same. At this point the workflow is interoperable and reproducible in general, and it can be registered in the OMERO Slurm Client's configuration file. To execute it from BIOMERO, we require two more files for which BIOMERO provides sensible defaults: the Slurm job script to run the container on Slurm with the right hardware parameters (e.g. memory, CPU, GPU and execution time) and the OMERO script to create a User Interface in OMERO and connect with the OMERO Slurm Client library. However, OMERO administrators are also encouraged to make their own custom scripts for either of these, as the Slurm job script defines the hardware requirements for the workflow (e.g., whether it needs a GPU or not) and the OMERO script defines a user interface and handles input/output from OMERO.

## Slurm job scripts

To schedule any workflows in Slurm, they require a Slurm job script. These scripts mainly describe the computational resources required (i.e. Memory, CPU and/or GPU), and the command to execute the workflow with the right parameters. Note that BIOMERO already provides default Slurm job scripts for every workflow configured. By default, no action will be required by the user or OMERO administrator to set up a Slurm job script, but we do allow providing custom scripts or customizing the default scripts from the OSC configuration file

In Figure 3, we show a detailed overview of all the layers of execution that are required (and automatically taken care of by our framework) to execute FAIR workflows from OMERO on HPC. The Slurm job script is what is actually submitted to the Slurm queue, and assigned and executed on the desired compute nodes. It is in the script that we call Singularity to run the Docker container on the assigned hardware. Another step we added to the Slurm job script is a conversion from ZARR to TIFF files, as all BIAFLOWS containers require TIFF input; yet we extract ZARR from OMERO. For efficiency, the conversions are also added in parallel to the Slurm queue, allowing the use of HPC infrastructure for speedier computation. Obviously, such conversion can be omitted when using workflows that natively execute on ZARR files.

One of the important factors not handled by the FAIR workflow setup from BIAFLOWS is the Slurm job script and its hardware parameters. Other workflow descriptor schemas like `boutiques`[17] do allow to define hardware parameters as part of the workflow metadata. BIOMERO provides sensible defaults and allows further customization of these parameters from its configuration file. We decided not to add these parameters to the OMERO User Interface to hide HPC implementation details from the end-user.

## Tutorials

We host a number of FAIR *Cytomine-0.1* workflows on GitHub, which we have updated and tested with BIOMERO. We have also included a tutorial for the creation of several of these containers, noted in the list below:

- CellPose[18], hosted at [https://github.com/TorecLuik/W_NucleiSegmentation-Cellpose](https://github.com/TorecLuik/W_NucleiSegmentation-Cellpose)
- CellProfiler for SpotCounting[19], hosted at [https://github.com/TorecLuik/W_SpotCounting-CellProfiler](https://github.com/TorecLuik/W_SpotCounting-CellProfiler)
    - Tutorial available at [https://nl-bioimaging.github.io/omero-slurm-client/tutorial_link.html#cellprofiler-tutorial](https://nl-bioimaging.github.io/omero-slurm-client/tutorial_link.html#cellprofiler-tutorial)
- CellExpansion, hosted at [https://github.com/TorecLuik/W_CellExpansion](https://github.com/TorecLuik/W_CellExpansion).
    - Tutorial available at [https://nl-bioimaging.github.io/omero-slurm-client/tutorial_link.html#cellexpansion-tutorial](https://nl-bioimaging.github.io/omero-slurm-client/tutorial_link.html#cellexpansion-tutorial)
- Spot counting script, hosted at [https://github.com/TorecLuik/W_CountMaskOverlap](https://github.com/TorecLuik/W_CountMaskOverlap).
    - Part of the same tutorial as CellExpansion.

To make it easier to get started with BIOMERO, we have also hosted two tutorials for setting up a simple Slurm cluster to connect your OMERO to:

- either on a local computer using Docker: https://nl-bioimaging.github.io/omero-slurm-client/tutorial_link.html#google-cloud-slurm-tutorial
- or using Google Cloud: https://nl-bioimaging.github.io/omero-slurm-client/tutorial_link.html#google-cloud-slurm-tutorial

## Discussion

In conclusion, our open-source framework paves the way for more widespread accessibility to FAIR bioimaging workflows, fostering collaboration and advancing the capabilities in bioimaging research. We promote best practices by funneling workflows in our BIOMERO into a FAIR shape, and by reusing workflows from Cytomine / BIAFLOWS, but we also enable a lot more users to execute these workflows by integrating with the OMERO web interface. We present a Python library enabling integration of OMERO and Slurm, allowing improved scalability and efficiency in bioimaging workflow execution and promoting best computation practices.

We want to note that our BIOMERO is modular, extensible and permissively open source, so while we have provided a first integration with OMERO using OMERO.scripts, other implementations are also possible and encouraged. Cytomine itself is also actively being developed at the time of writing, and will also improve upon the BIAFLOWS containers, so it is a well-suited bioimaging platform when not using OMERO.

Current limitations are mostly related to our workflow design being based on BIAFLOWS workflows: support for any generic container workflow is limited to only interpreting metadata values in the metadata descriptor JSON file that correspond to the `Cytomine-0.1` schema[20]. We also assume at this point that each workflow will accept an input folder of 8 bit/16-bit TIFF (2D) or single file OME-TIFF (C,Z,T) images. Both are current implementations based on BIAFLOWS workflows but are not inherent to our framework and more use-cases can and will be supported in the future. A third limitation is the OMERO.scripts interface. The two main blocking features are the static interface, disallowing user interactivity, and the absence of progress feedback: the user is only notified when the (long running) process is finished. A final limitation is that transferring large datasets (e.g. terabytes) to a remote Slurm cluster might be a costly operation by itself, depending on the connectivity between the servers.

## The next steps

In future versions, we will support more (generic) workflow descriptor languages, besides `cytomine-0.1`. We suggest supporting `boutiques` (on which `cytomine-0.1` is based) and/or the common workflow language (CWL)[21]. One of the major features that would be useful from such extended schemas is the description of required computational resources, which should be integrated into the generated Slurm job script.

We also plan to implement and support workflow management systems or languages, to chain modules together in a structured fashion, for example SnakeMake[22], NextFlow[23] or MLFlow[24] or even FRACTAL.

Moreover, we aim at supporting OME-NGFF[25] workflows, instead of being limited to TIFF input. This should be relatively straightforward since we already export images as ZARR. However, we are also aware that OME-NGFF is designed to be remotely accessed, reducing the data transfer costs significantly, so in the future we might only have to transfer a URI pointing to the original file (given that the Slurm cluster can access this endpoint too). This will reduce the strain on the framework and OMERO server.

Finally, we will develop a reactive user interface that allows easier access to the workflows, to replace the OMERO.scripts UI, and to make workflows more findable by integrating with existing workflow registries such as WorkflowHub.eu[26] or the BioImaging Zoo[27]. Within OMERO, a user should also be able to search for an existing workflow, or in other words, workflows should be more *findable and accessible*.

## Code availability

All original code for the OMERO Slurm Client has been deposited at Zenodo under the DOI 10.5281/zenodo.8409856 and is publicly available as of the date of publication at GitHub (https://github.com/NL-BioImaging/omero-slurm-client ) and PyPI (https://pypi.org/project/omero-slurm-client ), under the permissive Apache 2.0[28] license. We also provide the example BIOMERO Scripts (https://github.com/NL-BioImaging/biomero-scripts) repository allowing execution of 2D image-to-image workflows directly from the OMERO web interface script interface as described in this paper. These are in a separate repository on GitHub with the more restrictive copyleft GPL-2.0[29] license, as parts are copied from OME's GPL-2.0 licensed work, and can be installed directly on the OMERO.server to get started with BIOMERO, see the READMEs on GitHub for more details.

## Acknowledgments

This publication is part of the project NL-BioImaging-AM (with project number 184.036.012 of the National Roadmap research programme which is (partly) financed by the Dutch Research Council (NWO). Special thanks to our collaborators within the NL-BioImaging project for their valuable contributions. We also appreciate the support and insights provided by the bioimaging community and OME team on image.sc, which greatly enhanced the quality and scope of this research. Icons used in images are (adapted) from Material Design Icons (https://pictogrammers.com/library/mdi/ ) and OpenClipart (https://openclipart.org).

## Author Contributions

Conceptualization, TL and ER and RH and PK; Methodology, TL and PK and RH; Software, TL; Validation, RR; Writing – Original draft, TL; Writing – Review & Editing, PK and RH and ER and RR; Funding Acquisition, ER; Supervision, PK and RH

## Declaration of Interests

The authors declare no competing interests.

## Declaration of generative AI and AI-assisted technologies in the writing process

During the preparation of this work the author(s) used ChatGPT in order to improve readability and language of the text and generate drafts for paragraphs. After using this tool/service, the author(s) reviewed and edited the content as needed and take(s) full responsibility for the content of the publication.

## References

1. Allan, C., Burel, J.-M., Moore, J., Blackburn, C., Linkert, M., Loynton, S., MacDonald, D., Moore, W.J., Neves, C., Patterson, A., et al. (2012). OMERO: flexible, model-driven data management for experimental biology. Nat Methods *9*, 245–253. 10.1038/nmeth.1896.

2. Schindelin, J., Arganda-Carreras, I., Frise, E., Kaynig, V., Longair, M., Pietzsch, T., Preibisch, S., Rueden, C., Saalfeld, S., Schmid, B., et al. (2012). Fiji: an open-source platform for biological-image analysis. Nat Methods *9*, 676–682. 10.1038/nmeth.2019.

3. Bankhead, P., Loughrey, M.B., Fernández, J.A., Dombrowski, Y., McArt, D.G., Dunne, P.D., McQuaid, S., Gray, R.T., Murray, L.J., Coleman, H.G., et al. (2017). QuPath: Open source software for digital pathology image analysis. Sci Rep *7*, 16878. 10.1038/s41598-017-17204-5.

4. Slurm Workload Manager (2023). Wikipedia. https://en.wikipedia.org/w/index.php?title=Slurm_Workload_Manager&oldid=1153183551.

5. Marée, R., Rollus, L., Stévens, B., Hoyoux, R., Louppe, G., Vandaele, R., Begon, J.-M., Kainz, P., Geurts, P., and Wehenkel, L. (2016). Collaborative analysis of multi-gigapixel imaging data using Cytomine. Bioinformatics *32*, 1395–1401. 10.1093/bioinformatics/btw013.

6. Rubens, U., Mormont, R., Paavolainen, L., Bäcker, V., Pavie, B., Scholz, L.A., Michiels, G., Maška, M., Ünay, D., Ball, G., et al. (2020). BIAFLOWS: A Collaborative Framework to Reproducibly Deploy and Benchmark Bioimage Analysis Workflows. Patterns *1*, 100040. 10.1016/j.patter.2020.100040.

7. Seiffarth, J., Scherr, T., Wollenhaupt, B., Neumann, O., Scharr, H., Kohlheyer, D., Mikut, R., and Nöh, K. (2022). ObiWan-Microbi: OMERO-based integrated workflow for annotating microbes in the cloud. Preprint at bioRxiv, 10.1101/2022.08.01.502297 10.1101/2022.08.01.502297.

8. Overview | OMERO Plus | Glencoe Software, Inc. https://www.glencoesoftware.com/products/omeroplus/.

9. HCS | Image Data Solutions | Glencoe Software, Inc. https://www.glencoesoftware.com/solutions/hcs/.

10. Fractal Analytics Platform Fractal Analytics Platform. https://fractal-analytics-platform.github.io/.

11. Omero Slurm Client 10.5281/zenodo.8108214.

12. Welcome to Fabric! — Fabric documentation https://www.fabfile.org/.

13. Welcome to Omero Slurm Client's documentation! — Omero Slurm Client 1 documentation https://nl-bioimaging.github.io/omero-slurm-client/.

14. Luik, T. (2023). BIOMERO Scripts. https://github.com/NL-BioImaging/biomero-scripts.

15. omero-scripts/omero/export_scripts/Batch_Image_Export.py at develop · ome/omero-scripts GitHub. https://github.com/ome/omero-scripts/blob/develop/omero/export_scripts/Batch_Image_Export.py.

16. What is Secure File Copy (scp)? | SSH Academy https://www.ssh.com/academy/ssh/scp.

17. Glatard, T., Kiar, G., Aumentado-Armstrong, T., Beck, N., Bellec, P., Bernard, R., Bonnet, A., Camarasu-Pop, S., Cervenansky, F., Das, S., et al. (2017). Boutiques: a flexible framework for automated application integration in computing platforms. Preprint at arXiv, 10.48550/arXiv.1711.09713 10.48550/arXiv.1711.09713.

18. Stringer, C., Wang, T., Michaelos, M., and Pachitariu, M. (2021). Cellpose: a generalist algorithm for cellular segmentation. Nat Methods *18*, 100–106. 10.1038/s41592-020-01018-x.

19. cellprofiler-practical-NeuBIAS-Lisbon-2017/practical-handout.md at master · tischi/cellprofiler-practical-NeuBIAS-Lisbon-2017 GitHub. https://github.com/tischi/cellprofiler-practical-NeuBIAS-Lisbon-2017/blob/master/practical-handout.md.

20. Cytomine app JSON descriptor Reference | Cytomine ULiège R&D Documentation https://doc.uliege.cytomine.org/dev-guide/algorithms/descriptor-reference.

21. Crusoe, M.R., Abeln, S., Iosup, A., Amstutz, P., Chilton, J., Tijanić, N., Ménager, H., Soiland-Reyes, S., Gavrilović, B., Goble, C., et al. (2022). Methods included: standardizing computational reuse and portability with the Common Workflow Language. Commun. ACM *65*, 54–63. 10.1145/3486897.

22. Köster, J., and Rahmann, S. (2012). Snakemake—a scalable bioinformatics workflow engine. Bioinformatics *28*, 2520–2522. 10.1093/bioinformatics/bts480.

23. Di Tommaso, P., Chatzou, M., Floden, E.W., Barja, P.P., Palumbo, E., and Notredame, C. (2017). Nextflow enables reproducible computational workflows. Nat Biotechnol *35*, 316–319. 10.1038/nbt.3820.

24. MLflow - A platform for the machine learning lifecycle MLflow. https://mlflow.org/.

25. Moore, J., Allan, C., Besson, S., Burel, J.-M., Diel, E., Gault, D., Kozlowski, K., Lindner, D., Linkert, M., Manz, T., et al. (2021). OME-NGFF: a next-generation file format for expanding bioimaging data-access strategies. Nat Methods *18*, 1496–1498. 10.1038/s41592-021-01326-w.

26. Implementing FAIR Digital Objects in the EOSC-Life Workflow Collaboratory 10.5281/zenodo.4605654.

27. Ouyang, W., Beuttenmueller, F., Gómez-de-Mariscal, E., Pape, C., Burke, T., Garcia-López-de-Haro, C., Russell, C., Moya-Sans, L., de-la-Torre-Gutiérrez, C., Schmidt, D., et al. (2022). BioImage Model Zoo: A Community-Driven Resource for Accessible Deep Learning in BioImage Analysis. Preprint at bioRxiv, 10.1101/2022.06.07.495102 10.1101/2022.06.07.495102.

28. Apache License, Version 2.0 https://www.apache.org/licenses/LICENSE-2.0.

29. GNU General Public License v2.0 - GNU Project - Free Software Foundation https://www.gnu.org/licenses/old-licenses/gpl-2.0.html.

# Figure Legends

Figure 1 – BIOMERO manages connection between OMERO and HPC (local or cloud). At the center we have the open-source OMERO Slurm Client, a Python library. Users tell BIOMERO what workflow to run on what images, by selecting one of OMERO.scripts (A), available in the OMERO web interface. The OMERO script will communicate with the OMERO Slurm Client (B) to move data and track workflow execution. The OMERO Slurm Client downloads (if needed) and executes the required Docker containers (C), as jobs on the HPC (local or cloud). The job status and logs from the HPC executions are polled

through the OMERO Slurm Client (D) and, once the job is completed, results are stored back in OMERO (E) and communicated to the user via the OMERO web interface.

Figure 2 – Files required to run a workflow in BIOMERO. The underlying executable code needs to be a headless program (i.e. it does not require a graphical user interface). The input parameters should be described in a configuration file called descriptor.json, any preprocessing (or file reading) can be added to the wrapper.py, while the required environment should be defined (Python version, Python libraries, Fiji plugins, Matlab packages, OS, etc.) in the Dockerfile. These files should be published and versioned on GitHub (or another public online platform). From the Dockerfile a container image should be built, versioned and hosted on a public online container registry like DockerHub, to ensure reproducibility. At this point, anybody can pull the workflow to their local computer and run the workflow using Docker, Podman or Singularity or a dedicated workflow management system. Finally, to allow execution on HPC from OMERO, BIOMERO requires a Slurm job script that defines hardware parameters (we will generate a default one if none are provided) and an OMERO script that will generate the UI in OMERO.web (we provide a generic implementation of such an OMERO script that leverages the descriptor.json) and call on the OMERO Slurm Client. If these are all in place, the user can execute the workflow in OMERO via script UI.

Figure 3 - A workflow execution from BIOMERO will use a lot of automated layers of execution. It will start at the OMERO script 'SLURM Run Workflow' that will allow a user to select a workflow with its parameters (defined in descriptor.json on GitHub). Once selected, the OMERO script will export the selected input data (dataset, plate, or just a set of images) as ZARR files and wrap them up in a ZIP. The ZIP will be transferred to a Slurm node. The script will then order the workflow (preinstalled on Slurm) to be run. This will trigger a command on the Slurm's Linux shell to queue a new job based on a (preinstalled) job script and environment variables. The Slurm job script will first define the hardware configuration required for this workflow (GPU, CPU, etc.) but also a predefined execution duration and the location of the logfile. Next, it will trigger new Slurm jobs to convert all ZARR data to TIFF files and await their execution. Once completed, it will start a Singularity container for the workflow (preinstalled on Slurm from Docker) and await its completion. The container will have an installed environment (OS, libraries, etc.) defined in its Dockerfile, required for the workflow to execute properly. When run, the container will call upon an entrypoint script called wrapper.py (also defined in its Dockerfile and on GitHub). This wrapper script, based on BIAFLOWS, will read and validate the required parameters (defined in descriptor.json on GitHub) from the command line. Furthermore, the wrapper will handle any preprocessing required (e.g., cut images into smaller sizes), transfer data to a temporary folder before it will start a subprocess with the actual executable code (Fiji Macro, Matlab executable, Python or R script, CellProfiler pipeline, etc.). After execution is finished, generated output files will be moved to the 'out' folder, the temporary folder will be cleaned up, and Slurm will be informed of a successful execution. The OMERO script will poll the Slurm cluster for status updates during this execution and will start extracting output and log data when it finds the job COMPLETED. Once the data is back on the OMERO server, it will upload it into the OMERO system based on user-defined preferences (e.g., as images in a new dataset, or as a ZIP attachment for a project). Finally, all intermediate data will be cleaned up, and the OMERO.web UI will be informed of a successful script execution (with appropriate logging messages).

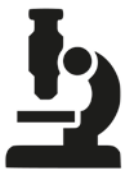
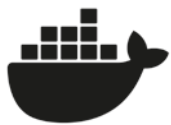
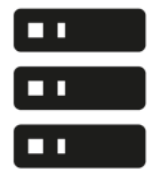
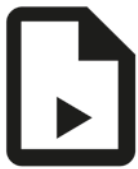
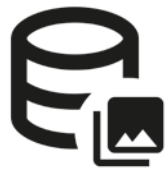
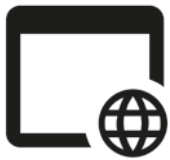
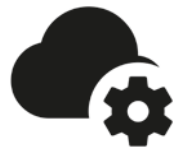

# Files required to run a headless workflow in the OMERO HPC Extension

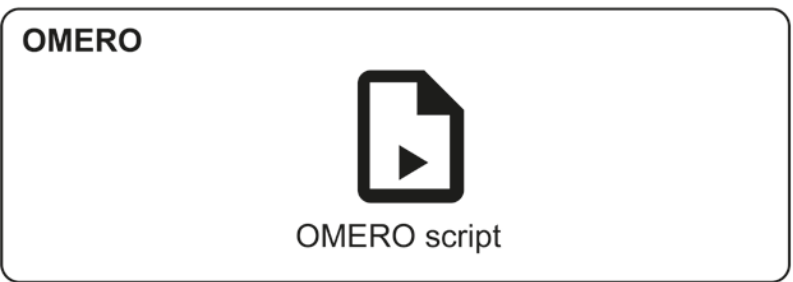

**OMERO**

OMERO script

The OMERO script will create a UI in OMERO web, given the metadata in the descriptor.json in GitHub

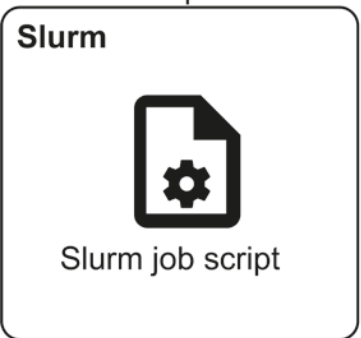

**Slurm**

Slurm job script

The Slurm job script calls the container with the right parameters and contains job and hardware parameters like CPU, GPU, memory and execution time

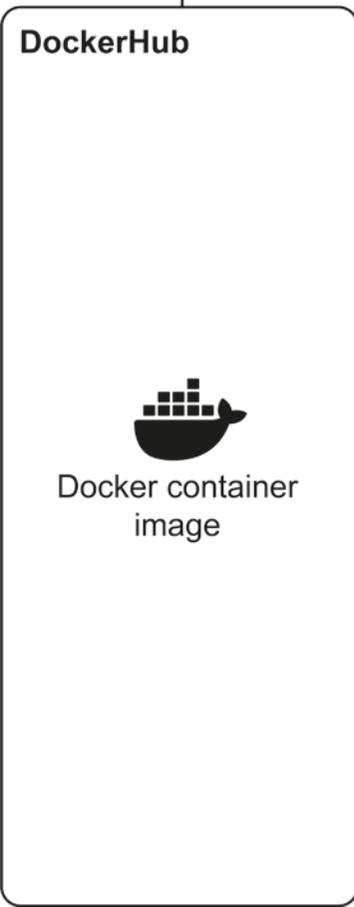

**DockerHub**

Docker container image

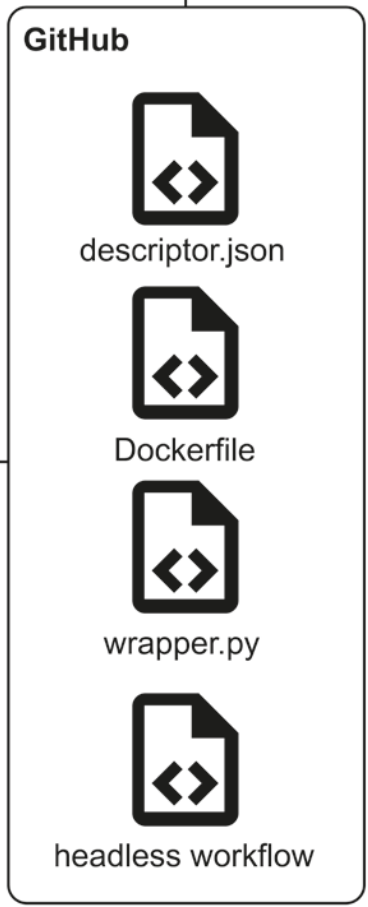

**GitHub**

descriptor.json

Dockerfile

wrapper.py

headless workflow

The container (image) is created, stored and versioned on a registry like DockerHub, from where it can be downloaded to Slurm

The files used to build the container (Dockerfile, the headless workflow and a wrapper.py) are stored and versioned on GitHub together with its metadata file called descriptor.json

# Automated layers of execution of a workflow from OMERO HPC Extension

**OMERO script**
- Create UI from workflow descriptor.json from GitHub
- (Optional) Validate/Setup remote Slurm environment
- Transfer input images from OMERO to Slurm (ZARR ZIP)

**Slurm shell**
- Queue new Slurm job based on Slurm job script
- parameters as environment variables
- Return Slurm jobid to OMERO script for polling

**Slurm job script**
- Define hardware parameters (MEM,CPU,GPU), job duration, logfile location
- (Optional) Convert input data, using new slurm jobs

**Slurm container**
- Environment from Dockerfile (OS, libraries, ...)
- Run wrapper.py with given parameters

**Slurm wrapper.py**
- Read commandline parameters as defined in descriptor.json
- Read data folder files, make tmp folder
- (Optional) Perform preprocessing

**Slurm subprocess**
- Execute headless workflow on tmp

- Copy tmp to out folder
- Cleanup tmp folder

- Log succesful execution

- Zip output folder

- (Ongoing) Poll Slurm job status until status changed
- If COMPLETED, retrieve logfile and zipped output folder
- If FAILED, retrieve logfile only
- Upload files to OMERO according to user instructions
- Cleanup remaining files on OMERO and Slurm

# Figure S1. Slurm folder structure, related to OMERO Slurm Client

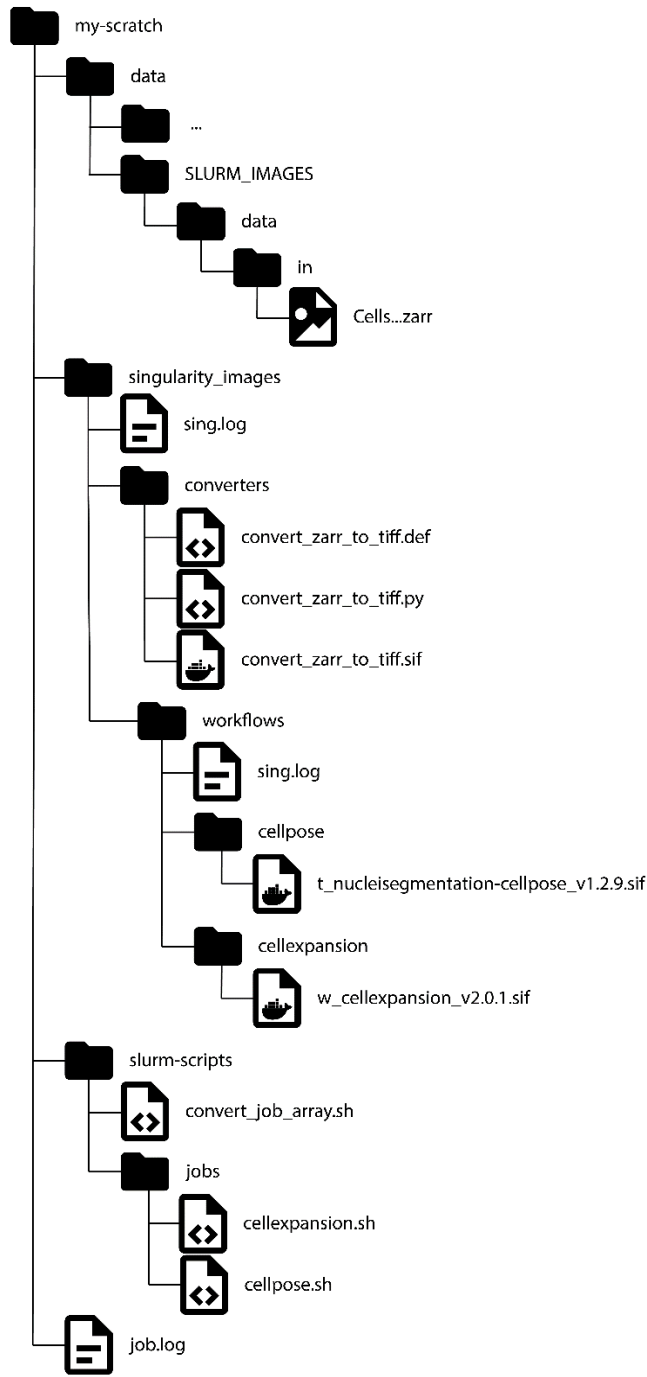

Figure S1 – Folder structure on Slurm nodes managed and created by OMERO Slurm Client (OSC). By design, all Slurm nodes share a storage in a folder ('my-scratch'), where logs of the running jobs are written. OSC populates this storage with 3 folders: "singularity_images", "slurm-scripts" and "data".

First and foremost, "singularity_images" contains all container images for all workflows and converters, downloaded in the Singularity Image Format (SIF). "slurm-scripts" contains a Slurm job script for each workflow in the jobs subfolder. This folder could be pulled from Git instead of generated. Finally, "data" will contain the actual datasets transferred from OMERO (as zip files). Data will be unpacked into a folder structure required for BIAFLOWS workflows, with subfolders in(put), out(put) and g(round)t(ruth); only 'in' is shown. All these elements are automatically set up and can be refreshed by OSC at any time when creating a connection.

# Note S2. OME-NGFF, related to OMERO HPC core scripts.

For workflows to be FAIR, it is important to have 1 type of data (image) input. The Cytomine-0.1 apps from BIAFLOWS all expect TIFF files as input (and output). However OME is working on the OME-NGFF[1] (ZARR) to be a future data standard for bioimaging, especially designed for image data analysis. To support both directions, we are exporting all selected images from OMERO to the ZARR format, using the OMERO-CLI-ZARR[2] app. On the Slurm cluster we run a converter from ZARR to TIFF format, using Slurm Array[3] capabilities to parallelize and queue the extra computation on the high-performance hardware. This way we support current Cytomine-0.1 apps with TIFF but also open the door to future apps using ZARR. This is an extendable design where future workflows could use other converters, or (preferably) work directly on the ZARR data.

# Supplemental References

1. Moore, J., Allan, C., Besson, S., Burel, J.-M., Diel, E., Gault, D., Kozlowski, K., Lindner, D., Linkert, M., Manz, T., et al. (2021). OME-NGFF: a next-generation file format for expanding bioimaging data-access strategies. Nat Methods *18*, 1496–1498. 10.1038/s41592-021-01326-w.

2. ome/omero-cli-zarr (2023). https://github.com/ome/omero-cli-zarr.

3. Slurm Workload Manager - Job Array Support https://slurm.schedmd.com/job_array.html.